\documentstyle[prb,aps,multicol,epsfig]{revtex}
\begin{document}
\title{Interlayer exchange coupling in M/N/M multilayer } 
\author{S.Mathi Jaya$^{*}$, M.C.Valsakumar$^{*}$ and W.Nolting$^{+}$}
\address{$^{*}$Materials Science Division, Indira Gandhi Centre for Atomic Research,
Kalpakkam 603102, India \\ 
$^{+}$Festk\"orpertheorie, Institut f\"ur Physik, Humboldt Universit\"at zu Berlin, Invaliden Strasse \ 110, 10115 Berlin, Germany}
\maketitle
\begin{abstract}
    The interlayer exchange coupling (IEC) of  two local moment ferromagnetic layers 
separated by a non-magnetic 
spacer layer (M/N/M multilayer) is studied using the  modified RKKY method along with the s-f model. The IEC 
exhibits oscillatory behaviour with respect to the spacer layer thickness and it oscillates 
between ferro- and antiferromagnetic configurations.  The conventional RKKY method is   also 
used to obtain  the IEC and the results are compared with those  obtained from the 
modified RKKY method which incorporates the electron correlation effects. We find significant 
correlation effects on the IEC and in fact the correlations alter the nature  and 
magnitude of the magnetic coupling. Hence  this study  points out the importance of the inclusion of 
correlation effects in the understanding of the IEC in multilayer systems with local moment 
ferromagnetic sub-layers and  to offer satisfactory explanation for the experimental results.   
\end{abstract}
\begin{multicols}{2}
\section{Introduction}
     The indirect exchange interaction between two ferromagnetic layers separated by a non-magnetic 
spacer layer exhibits oscillatory behaviour with respect to the spacer layer thickness and this 
has been observed  experimentally in many magnetic multilayer systems [1-6]. The interpretation  
of these experimental observations are provided by methods based on the RKKY interaction [7,8],  Hubbard 
type model [9], s-f exchange model [10] and ab-initio total energy calculations [11-13].  
Naive application of the RKKY method (assuming spherical Fermi surface and uniform distribution of spins in the 
ferromagnetic layers) leads to small oscillation periods [7]. Hence Bruno and Chappert [8]  extended the 
general theory of the RKKY method and got satisfactory oscillation periods for Co/Cu/Co and Fe/Cu/Fe  
multilayers considering the topological properties of the Fermi surface of the spacer layer and 
the moment distribution within the ferromagnetic layers. Using a Hubbard like one band  
model [9],  Edwards et. al.  showed that  the  exchange coupling  exhibits long period oscillations
with respect to the spacer layer thickness   for certain positions of the Fermi level.
Urbaniak-Kucharczyk used the s-f exchange model to study the interlayer coupling  expressing the effective 
exchange integrals in terms of the electron susceptibility [10].   
Convincing results for the fcc-(111),(100) and (110) spacers are found. Ab-initio total energy methods 
 are restricted to small spacer layer thickness as they become very difficult when the spacer 
thickness  is   large.  
 An extensive discussion of the 
general theory of the interlayer exchange is given by Bruno [14] and the illustration of the theory for the 
case of Co/Cu/Co multilayer is also presented by him.    The interlayer exchange coupling 
is found to have significant influence  on 
the magnetic properties of the ferromagnetic sublayers.  Ney et. al. found an oscillating behaviour 
in the $T_{c}$ of Co/Cu/Ni trilayers with respect to the spacer layer thickness [1] and a 
theoretical analysis of this effect based on the Hubbard model is  reported by Wu et.al. [15].   \\

     Our interest is to study  the  exchange coupling of the ferromagnetic (M) layers in 
 M/N/M multilayers with M being  a local moment 
ferromagnetic metal and N being  a nonmagnetic metal. In the  local moment metals, the magnetic 
properties are dominated by the intra-atomic exchange interaction acting between the conduction electrons 
and the local moments (s-f exchange) and the magnetic coupling between the local moments situated 
at the lattice sites is  mediated by the conduction electrons.  
The above mentioned correlation effects are properly taken into account in the s-f model which is
 often referred to 
as the ferromagnetic Kondo lattice model in the recent liturature   and 
the modified RKKY method [16,17] describes the exchange coupling of the localised spins mediated by the 
correlated conduction electrons. Thus this approach leads to the  self-consistent evaluation of  the  exchange 
integrals acting  between the local moments situated at the lattice sites   and 
hence we use this s-f model along with  the modified RKKY method to study 
the interlayer exchange coupling in M/N/M multilayers with M being a local moment metal.
The  starting Hamiltonians of our method and the method of Urbaniak-Kucharczyk [10] are similar but  the
 approaches are completely different (see sect.V). Further, our method and the  Urbaniak-Kucharczyk's method are different 
from the Bruno's method [14]  in the sense that in the  Bruno's method, the IEC is derived from the interference 
effects of the electron waves whereas in our method and the  Urbaniak-Kucharczyk's method, the IEC is 
derived using the indirect coupling of the localised spins mediated by the conduction electrons. 
In our method the full Green function of the $s-f$ system is 
evaluated and the interlayer exchange coupling (IEC) is studied using the above said modified RKKY method. 
In the present work, we have carried out our study for  a model system with only one conduction band per 
layer and we hope that 
this study will form the basis for the  calculation for a real system. The influence of the interlayer 
exchange coupling on the magnetic properties of the ferromagnetic sublayer can also be studied using 
this method. However this study  is planned for a future paper. In this paper we present our results on the interlayer 
exchange coupling in the M/N/M multilayer at T=0K with different thicknesses of the non-magnetic spacer 
layer.  We have further studied the influence of the conduction electron concentration on the IEC. We present 
our results in the following sections along with a brief discussion of the s-f model and the modified RKKY 
method.


\section{s-f model for the ferromagnetic  films}
    In this section we shall present a brief discussion of the s-f model. We shall present the 
theory for an 'n' layer film and the theory can then easily be adapted for the required M/N/M multilayer 
geometry by assuming that there are no localised spins in the non-magnetic spacer layers. 
   The theory and the mathematical formulation of the $s-f$ model are described in many of the 
earlier publications [16,17]. Hence we will not present those  details here. 
However, for the sake of completion, we  will present a brief summary of the model.  
    We will consider a ferromagnetic film with $n$ layers. The film is characterised by a two 
dimensional Bravais lattice vector having an 'n' atom basis. The 'n' atom basis corresponds to 
the 'n' layers of the  film. A lattice vector of the film may be  given as 
$$ {\bf R}_{i\alpha} = {\bf R}_{i} + {\bf r}_{\alpha} $$
${\bf R}_{i}$ is the two dimensional Bravais lattice vector and  ${\bf r}_{\alpha}$ is the
basis vector.  
At each site of the film  a localised spin 'S' is present and  
the film is  assumed to have only one  conduction band per layer.  The electrons in the conduction band 
are exchange coupled to the local moments. This situation is well described by the $s-f$ model,  
 the model Hamiltonian of which is given as 
$$H  = \sum_{{ij,\sigma \atop \alpha\beta}}t^{\alpha\beta}_{ij}c_{i\alpha\sigma}^{\dagger}
c_{j\beta\sigma}  
    -\jmath \sum_{j,\alpha}{{\bf S}_{j\alpha}}\cdot {{\bf \sigma}_{j\alpha}} $$ 
The first term describes the conduction electrons and the second term represents the interaction 
of the conduction electrons with the local moments. i,j represent the site indices of the two dimensional 
lattice and $\alpha$, $\beta$ represent the layer indices.   $t_{ij}^{\alpha\beta}$ is the hopping integral 
and $\jmath$ is the $s-f$ exchange coupling strength. 
All the information concerning the electronic structure and magnetic properties of the system 
described by the above said Hamiltonian can be obtained from the retarded single electron Green 
function 
$$G^{\alpha\beta}_{ij\sigma} =\  \ll c_{i\alpha\sigma} ; c_{j\beta\sigma}^{\dagger} \gg_{E} \ = $$ 
             $$  -i\int_{0}^{\infty}dt e^{\frac{i}{\hbar}Et} < [c_{i\alpha\sigma}(t),c_{j\beta\sigma}^{\dagger}(0)]_{+} >  $$
Evaluation of this Green function is proceeded by the equation of motion  method. The equation of motion 
is  written as 
$$\sum_{r\gamma}(E\delta_{ir}-T^{\alpha\gamma}_{ir})G^{\gamma\beta}_{rj\sigma}(E) = 
\hbar\delta_{ij}\delta_{\alpha\beta} 
 +\ll[c_{i\alpha\sigma},H_{sf}]_{-} ; c_{j\beta\sigma}^{\dagger}\gg_{E} $$

Defining  the selfenergy $M_{ij\sigma}(E)$ as  
$$\ll [c_{i\alpha\sigma},H_{sf}]_{-} ; c_{j\beta\sigma}^{\dagger}\gg_{E} = 
\sum_{r\gamma}M^{\alpha\gamma}_{ir\sigma}(E)G^{\gamma\beta}_{rj\sigma}(E) $$
and making a Fourier transformation (with respect to the spatial variables) 
 of the equation of motion  leads to
$$\sum_{\gamma}(E\delta_{\alpha\gamma}-\varepsilon_{\alpha\gamma}({\bf k})-
M^{\alpha\gamma}_{{\bf k}\sigma}(E))G^{\gamma\beta}_{{\bf k}\sigma}(E) = 
\hbar\delta_{\alpha\beta} $$

In the matrix form the above equation may be written as 
		$$\hat G_{{\bf k}\sigma}(E) = \frac{\hbar} {
	   [E\hat I-\hat \varepsilon({\bf k}) -\hat M_{{\bf k}\sigma}(E)] }$$

$\hat \varepsilon({\bf k})$ is the  Bloch energy matrix  and for an 'n' layer film it 
will take the form 

$$ \hat\varepsilon({\bf k}) = \left (\matrix {\varepsilon^{11}({\bf k}) & 
\varepsilon^{12}({\bf k}) & \cdot\cdot\cdot\cdot & \varepsilon^{1n}({\bf k}) \cr 
 \varepsilon^{21}({\bf k}) & \varepsilon^{22}({\bf k}) & \cdot\cdot\cdot\cdot & 
\varepsilon^{2n}({\bf k})  \cr 
{\cdot\atop\cdot}\atop{\cdot} & {\cdot\atop\cdot}\atop{\cdot} & 
{\cdot\atop\cdot}\atop{\cdot} & {\cdot\atop\cdot}\atop{\cdot}  \cr 
\varepsilon^{n1}({\bf k}) & \varepsilon^{n2}({\bf k}) & \cdot\cdot\cdot\cdot 
 & \varepsilon^{nn}({\bf k}) \cr } \right ) $$  \\
The diagonal elements are the intralayer Bloch energies and the off-diagonal elements are 
the interlayer Bloch energies. \\

$\hat M_{\sigma}(E)$ is the  self-energy matrix and it takes the form \\
$$ \hat M_{\sigma}(E) = \left (\matrix { M^{11}_{\sigma}(E) & M^{12}_{\sigma}(E)
 & \cdot\cdot\cdot\cdot &  M^{1n}_{\sigma}(E)  \cr 
  M^{21}_{\sigma}(E) & M^{22}_{\sigma}(E)  & \cdot\cdot\cdot\cdot &  M^{2n}_{\sigma}(E) \cr 
{\cdot\atop\cdot}\atop{\cdot} & {\cdot\atop\cdot}\atop{\cdot} & 
{\cdot\atop\cdot}\atop{\cdot} & {\cdot\atop\cdot}\atop{\cdot} \cr
M^{n1}_{\sigma}(E) & M^{n2}_{\sigma}(E) & \cdot\cdot\cdot\cdot & 
M^{nn}_{\sigma}(E) \cr } \right ) $$  \\
The diagonal elements are the intralayer self-energies and the off-diagonal elements are 
the interlayer self-energies. \\

The layer dependent spectral density may now be obtained from the Green function matrix as  
$$ S_{{\bf k}\sigma}^{\alpha\alpha}(E) = - \frac{1}{\pi}Im G_{{\bf k}\sigma}^{\alpha\alpha}(E) $$ 
The layer dependent quasiparticle density of states (QDOS) is 
$$ \rho_{\sigma}^{\alpha\alpha}(E) = \sum_{{\bf k}} S_{{\bf k}\sigma}^{\alpha\alpha}(E) $$ \\

The  evaluation of the self-energy is   described in 
many of the earlier publications [18-20]. Once the self-energy is calculated, the 
 matrix Green function of the film can immediately be obtained.   

\section{Modified RKKY method } 
    The modified RKKY method essentially aims at the evaluation of  the indirect 
exchange coupling between two localised 
moments  mediated by the conduction electrons which are coupled to the localised spins 
situated at the lattice sites  through 
the $s-f$ exchange acting between them  and the localised spins. In order to obtain this effective  exchange 
interaction between the localised spins, the $s-f$ interaction is mapped to an effective Heisenberg  
Hamiltonian by averaging out the conduction electron degrees of freedom [17]. 
 This procedure allows us  to introduce an effective exchange coupling between the 
localised moments as a functional of the conduction electron self-energy. The details of the method may again be 
found in the earlier publications [17] and hence we will not elaborate the method here. 
However for the sake of completion, we will  quote here the required results which are used in the calculations. 
The exchange integral acting between two localised moments is given by this method as  
$$ J_{ij}^{\alpha\beta} = \sum_{\bf q}J^{\alpha\beta}({\bf q})
 e^{-i{\bf q}\cdot({\bf R}_{i\alpha}-{\bf R}_{j\beta})} $$ 
where $J^{\alpha\beta}({\bf q})$ is given  as  
$$J^{\alpha\beta}({\bf q}) = \frac{1}{8\pi}\jmath^{2}\sum_{\sigma}Im\int_{-\infty}^{\infty}dEf_{-}(E)
\frac{1}{N\hbar}\sum_{\bf k}A_{{\bf k},{\bf k+q}}^{\sigma\sigma,\alpha\beta}(E) $$  and    
 $$\hbar A_{{\bf k},{\bf k+q}}^{\sigma\sigma,\alpha\beta}(E)= (\hat G_{\bf k}^{(0)}(E) 
\hat G_{\bf k+q\sigma}(E) )^{\alpha\beta} +
(\hat G_{\bf k+q}^{(0)}(E) \hat G_{\bf k\sigma}(E) )^{\alpha\beta} $$ 
$\hat G_{\bf k}^{(0)\alpha\beta} $ is the free electron Green function and 
$\hat G_{\bf k\sigma}^{\alpha\beta}(E)$ is the already 
defined Green function of the $s-f$ system. $f_{-}(E)$ is the Fermi function. 
The ${\bf q}$ dependent exchange integrals, thus
can also be written in a matrix form for an 'n' layer film as 
$$ \hat J({\bf q}) = \left (\matrix {J^{11}({\bf q}) & 
J^{12}({\bf q}) & \cdot\cdot\cdot\cdot & J^{1n}({\bf q})  \cr 
 J^{21}({\bf q}) & J^{22}({\bf q}) & \cdot\cdot\cdot\cdot & 
J^{2n}({\bf q})  \cr 
{\cdot\atop\cdot}\atop{\cdot} & {\cdot\atop\cdot}\atop{\cdot} & 
{\cdot\atop\cdot}\atop{\cdot} & {\cdot\atop\cdot}\atop{\cdot}  \cr 
J^{n1}({\bf q}) & J^{n2}({\bf q}) & \cdot\cdot\cdot\cdot 
 & J^{nn}({\bf q}) \cr } \right ) $$ 
As the exchange integrals are dependent on the Green function of the $s-f$ system, it is 
obvious that they are dependent on the  electron self-energy. \\ 

     It is possible to obtain the conventional RKKY interaction  from this formalism  through the 
first-order approximation 
$$ G_{\bf k\sigma}^{\alpha\beta}(E) \rightarrow  G_{\bf k\sigma}^{(0)\alpha\beta}(E) $$ 
$$ G_{\bf k+q\sigma}^{\alpha\beta}(E) \rightarrow  G_{\bf k+q\sigma}^{(0)\alpha\beta}(E) $$ 
In  this case $J^{\alpha\beta}({\bf q})$ will be identical to the well-known RKKY expression. 

\section{Interlayer exchange coupling } 
    Our primary interest is to evaluate the indirect exchange coupling between two local moment ferromagnetic 
layers separated by a nonmagnetic spacer layer  including the effects of the above mentioned $s-f$ exchange 
acting between the conduction electrons and the localised spins of the ferromagnetic layers.
Hence we have made use of the modified RKKY method to evaluate the required exchange 
integrals. 
The matrix Green function of the 'n' layer film can easily be converted to describe the 
M/N/M multilayer by assuming that there are no localised spins in the spacer layers and hence 
the Green functions of the spacer layers will be just the free electron Green functions. The matrix 
Green function of the M/N/M multilayer thus can be generated and it can subsequently be used to 
obtain the exchange integral matrix $\hat J({\bf q})$ discussed in the previous section. The exchange interaction 
acting between two  moments located in the bottom and topmost ferromagnetic sub-layers ($J_{ij})$  of the multilayer
can now be calculated from $\hat J({\bf q})$. 
Summation  over all the exchange integrals $(J_{ij})$  
acting between the  localised moments situated at the lattice sites of these two ferromagnetic layers 
will yield the interlayer exchange coupling. 
In all our calculations, we assumed the geometry of two ferromagnetic monolayers separated by a spacer 
layer and the spacer layer thickness was varied upto thirty monolayers. The geometry of the multilayer 
with a spacer thickness of five monolayers is shown in Fig.1.  

The Bravais lattice of each 
layer is assumed to be a squre lattice and the  nearest neighbor intralayer and interlayer 
electron hopping is assumed to occur in the multilayer. The magnitude of the hopping is assumed 
to be the same in all the layers.  
 The value of the $s-f$ coupling constant $\jmath$ is assumed to 
be 0.2 eV and the magnitude of the localised spin is taken as 7/2.  
The electron self-energy is
 actually dependent on the band occupation and temperature and 
hence the exchange integrals will also be dependent on the band occupation and temperature. We have
studied the influence of the band occupation on the exchange integrals keeping the same 
 band occupation in all the layers and the influence of temperature on the interlayer exchange 
coupling is planned for a forthcoming paper. 
    We have further evaluated the interlayer coupling using the conventional RKKY method and the 
results of our calculations are discussed in the following section. 
\vspace{1.5cm}
\begin{figure}[htbp]
   \centerline{\epsfxsize=6cm \hspace{.02cm} \epsffile{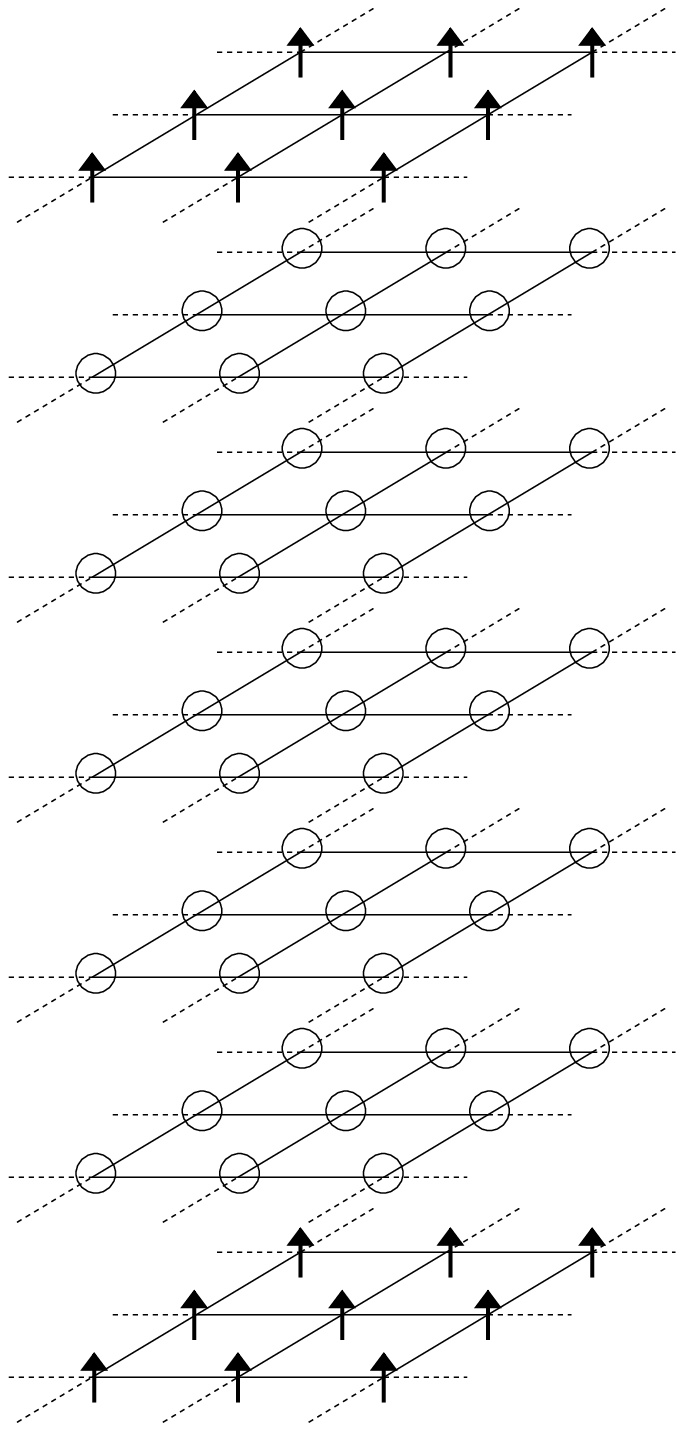}}
\vspace{0.0cm}
{Fig.1 Geometry of the M/N/M multilayer with a spacer thickness of 5 monolayers }
\end{figure}

\section{Discussion of results }
     The calculated interlayer exchange coupling  (IEC) using the modified RKKY method 
at two different band occupations (n=0.2, 0.8)  
are presented in Figs. 2 and 3. The IEC oscillates between  ferro- and antiferromagnetic 
configurations. This behaviour is in agreement with that of the trend seen in many experimental
works. As it is a model calculation, the oscillation periods and nature of the magnetic 
coupling with respect to the spacer layer thickness cannot be compared quantitatively with the experimental 
works. Hence our next goal is to extend the calculations to a real material film where the 
calculational complexity  however will be enormous. 
\vspace{-2.0cm}
\begin{figure}[htbp]
   \centerline{\epsfxsize=12cm \hspace{2cm} \epsffile{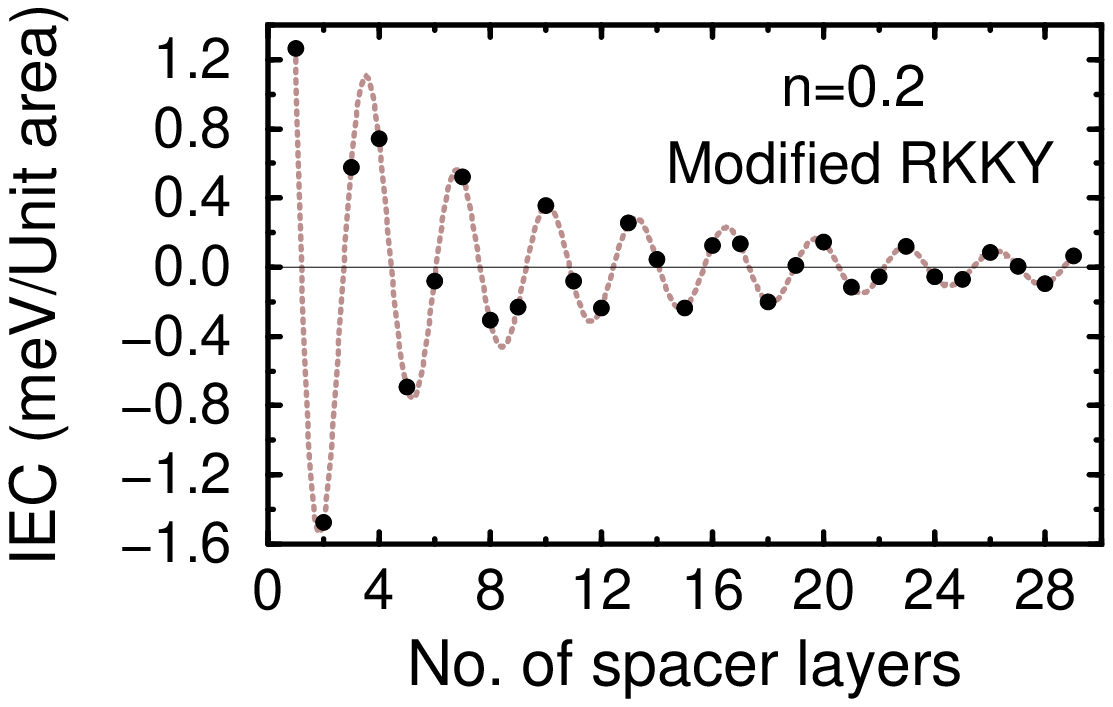}}
\vspace{-0.5cm}
{Fig.2 The interlayer exchange coupling (IEC) obtained from  the modified RKKY method at various 
values of the spacer thickness at a band occupation of 0.2 (T=0K) }
\end{figure}
\vspace{-3.0cm}

\begin{figure}[htbp]
   \centerline{\epsfxsize=12cm \hspace{2cm} \epsffile{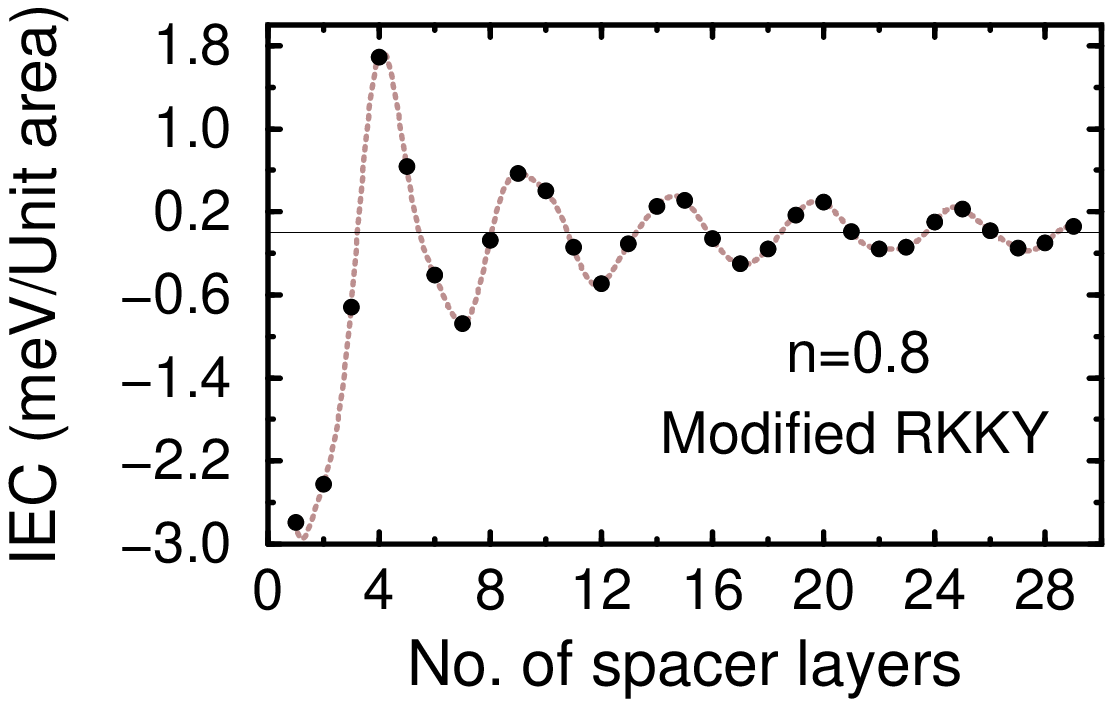}}
\vspace{-0.5cm}
{Fig.3 The interlayer exchange coupling (IEC) obtained from  the modified RKKY method at various 
values of the spacer thickness at a band occupation of 0.8 (T=0K) }
\end{figure}

Our interest in this study is to point out
the importance of electron correlation effects (self-energy) in deciding the nature of the magnetic
coupling of the ferromagnetic layers with respect to the spacer layer thickness and in 
influencing the oscillation periods. Hence we carried out the calculations  using the 
conventional RKKY method also. The results obtained using the conventional RKKY method  
 for the same band occupations  are presented in Figs. 4 and 5.
It may be seen from the figures that the influence of correlation on the magnetic coupling is 
significant. When the spacer consists of  one monolayer, the coupling is antiferromagnetic for the 
uncorrelated film (Fig.4) whereas it is ferromagnetic for the correlated film (Fig.2). 
The behaviour of the magnetic coupling at larger spacer layer thicknesses also show 
significant differences between the uncorrelated and correlated films. The magnitude of the 
coupling strength is also drastically modified because of the correlation
effects and this may also be seen from the figures. Thus our studies clearly demonstrates the influence of the 
correlation effects on the exchange coupling of the ferromagnetic  layers.
\vspace{-3.0cm}
\begin{figure}[htbp]
   \centerline{\epsfxsize=12cm \hspace{2cm} \epsffile{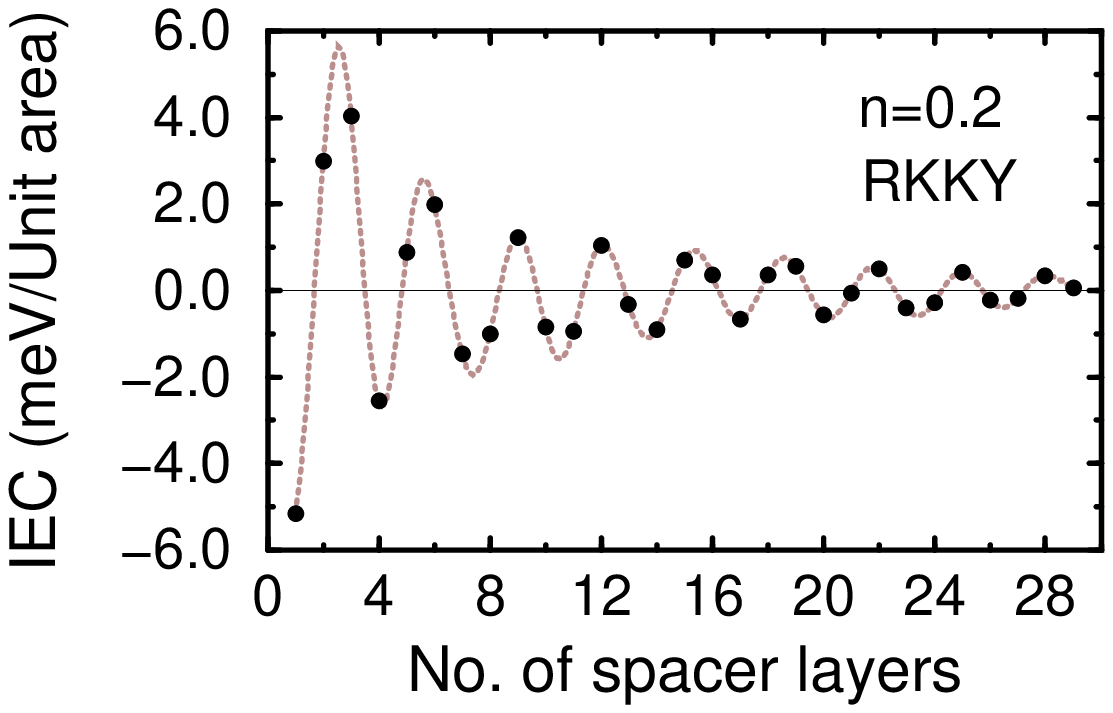}}
\vspace{-0.5cm}
{Fig.4 The interlayer exchange coupling (IEC) obtained from  the conventional RKKY method at various 
values of the spacer thickness at a band occupation of 0.2 (T=0K) }
\end{figure}
\vspace{-3.0cm}
\begin{figure}[htbp]
   \centerline{\epsfxsize=12cm \hspace{2cm} \epsffile{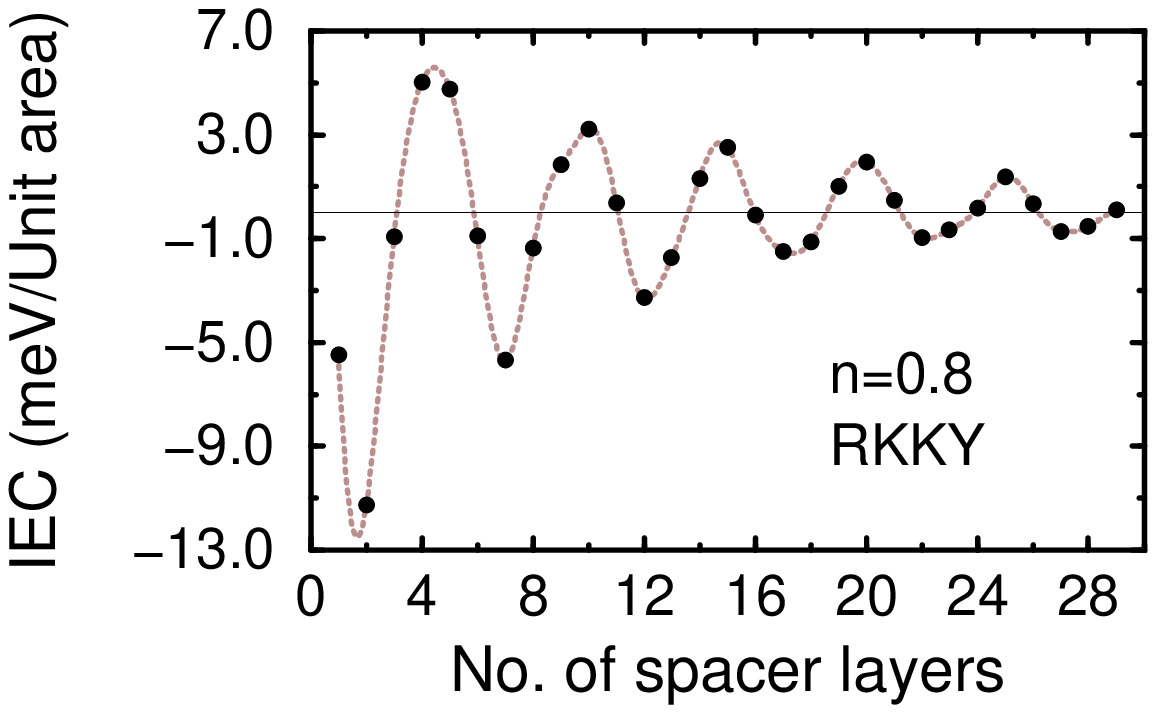}}
\vspace{-0.5cm}
{Fig.5 The interlayer exchange coupling (IEC) obtained from  the conventional RKKY method at various 
values of the spacer thickness at a band occupation of 0.8 (T=0K) }
\end{figure}
 
     In order to study the evolution of the correlation effects on the interlayer coupling, 
we have further calculated the IEC  as a function of the  $s-f$ exchange
coupling strength ($\jmath$) for the multilayer with a single spacer layer using the modified RKKY as well
as the conventional RKKY methods. The results are plotted in Fig.6. It may be seen from the 
figure that, both the methods predict 
antiferromagnetic exchange coupling  at small values of $\jmath$. However, after a critical value
 of $\jmath$,
the modified RKKY method predicts a transition of the exchange coupling to ferromagnetic nature, 
whereas the conventional RKKY method predicts that the antiferromagnetic coupling is retained 
for all values of  $\jmath$.  The effect of correlation on the IEC is thus clearly demonstrated in Fig.6.
The modified RKKY method further predicts that the IEC gets saturated after certain value of J, whereas in the 
case of conventional RKKY method, the IEC can never reach saturation.  
As the modified RKKY method uses the full Green function of the system which properly  takes into account of the 
$s-f$ interaction, it is obvious that these calculations are  much more reliable than the conventional
RKKY method. On the other hand, in the method of  Urbaniak-Kucharczyk the effective exchange integrals 
are expressed in terms of the electron susceptibility and the $s-f$ interaction strength $\jmath$. The expression 
for the effective exchange integrals is quadratic in $\jmath$ and hence the behaviour of the IEC seen in Fig.6 
is not abtainable from his method.  
The  IEC is found to be sensitive to the   band occupation  also and  
it may  be seen from the Figs. 2, 3 and  4, 5. The band occupation also alters the nature and magnitude of 
the IEC.
\vspace{-3.0cm}
\begin{figure}[htbp]
   \centerline{\epsfxsize=12cm \hspace{2cm} \epsffile{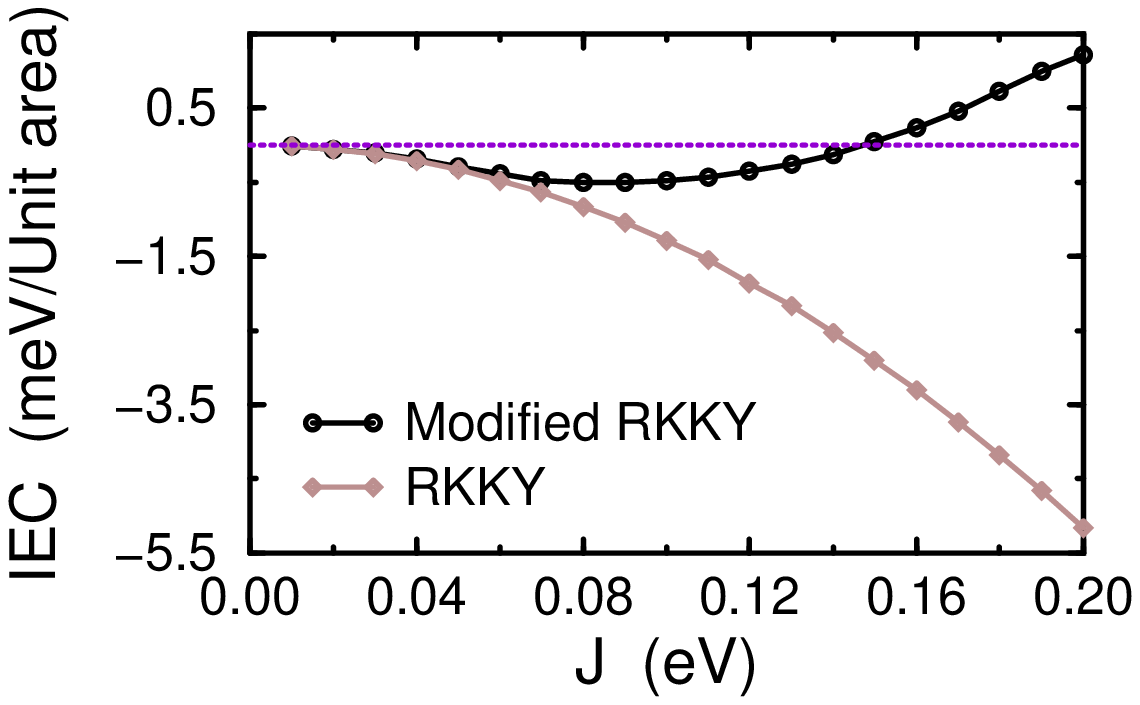}}
\vspace{-0.5cm}
{Fig.6 Variation of the  interlayer exchange coupling (IEC) with respect to the $s-f$ exchange coupling 
strength ($\jmath$) for the M/N/M multilayer with a monolayer spacer (T=0K) }
\end{figure}

 In order to obtain the oscillation periods, we performed a Fourier analysis of our data. 
The dotted lines shown in Figs.2-5 are the Fourier fit to our data. The Fourier analysis of
 our data corresponding to the  modified RKKY method 
 revealed that the oscillation period at 
n=0.2 is  3.22 whereas as it is 4.83 when n=0.8. Thus we found only a single period oscillation 
in the IEC.  The oscillation periods are almost the same for 
the RKKY data also. It may be seen from Figs.2-5 that there is a overall sign change of the decaying 
wave from the  modified RKKY to conventional RKKY for n=0.2, whereas it is absent for n=0.8. The magnitude 
of the oscillation is very much reduced in the modified RKKY compared to that of the conventional RKKY.

\section{Conclusions }
 We have evaluated the interlayer exchange coupling of  two local moment ferromagnetic sublayers in a M/N/M 
multilayer at different spacer thicknesses.  The calculations were carried out using the  $s-f$ model and the modified RKKY  method
  which provide a 
 self-consistent description of the exchange coupling between the moments situated at the lattice sites of the two ferromagnetic 
sublayers. 
 The IEC is found to have an oscillating behaviour with respect to the 
spacer layer thickness and it oscillates between ferro- and antiferromagnetic configurations. Inorder to demonstrate the influence 
of correlation effects on the IEC, we have further evaluated it using the conventional RKKY method too. 
We find significant influences of the  electron correlation effects on the IEC and the correlation effects are found to change the 
nature and magnitude of the exchange coupling. 
The variation of the IEC with respect to the band occupation is also studied using both the methods and the band 
occupation also found to alter the nature and magnitude of the IEC. The oscillation period of the IEC  is also found to 
depend on the band occupation.   Our method is very much sophisticated  than 
the conventional RKKY method and hence a calculation for real local moment multilayers can be expected  give 
reliable and interesting results.

\end{multicols}

\end{document}